\documentclass[twocolumn,11pt]{aastex62}
\usepackage{booktabs}
\usepackage{amsmath}
\usepackage{graphicx}
\usepackage{CJK}
\usepackage{subfigure}
\usepackage{threeparttable}

\usepackage[switch]{lineno} 

\long\def\symbolfootnote[#1]#2{\begingroup%
\def\thefootnote{\fnsymbol{footnote}}\footnote[#1]{#2}\endgroup}

\newcommand{\beq}{\begin{equation}}
\newcommand{\eeq}{\end{equation}}
\newcommand{\bea}{\begin{eqnarray}}
\newcommand{\eea}{\end{eqnarray}}

\begin{document}


\title{\Large \bf Quasi-periodic oscillation in short gamma-ray bursts from black hole-neutron star mergers}

\author{Yan Li (李彦)}\thanks{liyan287@mail2.sysu.edu.cn, now at Guilin University of Technology}
\affiliation{School of Physics and Astronomy, Sun Yat-Sen University, Zhuhai, 519082, P. R. China}
\affiliation{CSST Science Center for the Guangdong-Hongkong-Macau Greater Bay Area, Sun Yat-Sen University, Zhuhai, 519082, P. R. China}
\author{Rong-Feng Shen (申荣锋)}\thanks{shenrf3@mail.sysu.edu.cn}
\affiliation{School of Physics and Astronomy, Sun Yat-Sen University, Zhuhai, 519082, P. R. China}
\affiliation{CSST Science Center for the Guangdong-Hongkong-Macau Greater Bay Area, Sun Yat-Sen University, Zhuhai, 519082, P. R. China}
\author[0000-0003-4111-5958]{Bin-Bin Zhang (张彬彬)}\thanks{bbzhang@nju.edu.cn}
\affiliation{School of Astronomy and Space Science, Nanjing University, Nanjing 210093, China}
\affiliation{Key Laboratory of Modern Astronomy and Astrophysics (Nanjing University), Ministry of Education, China}
\affiliation{Purple Mountain Observatory, Chinese Academy of Sciences, Nanjing 210023, China}

\begin{CJK*}{UTF8}{gbsn}

\begin{abstract}

Short-duration gamma-ray bursts (sGRBs) are commonly attributed to the mergers of double neutron stars (NSs) or the mergers of a neutron star with a black hole (BH). While the former scenario was confirmed by the event GW170817, the latter remains elusive. Here, we consider the latter scenario in which, a NS is tidally disrupted by a fast spinning low-mass BH and the accretion onto the BH launches a relativistic jet and hence produces a sGRB. The merging binary's orbit is likely misaligned with the BH's spin. Hence, the Lense-Thirring precession around the BH may cause a hyper-accreting thick disk to precess in a solid-body manner. We propose that a jet, initially aligned with the BH spin, is deflected and collimated by the wind from the disk, therefore being forced to precess along with the disk. This would result in a quasi-periodic oscillation or modulation in the gamma-ray light curve of the sGRB, with a quasi-period of $\sim 0.01-0.1$ s. The appearance of the modulation may be delayed respective to the triggering of the light curve. This feature, unique to the BH-NS merger, may have already revealed itself in a few observed sGRBs (such as GRB 130310A), and it carries the spin-obit orientation information of the merging system. Identification of this feature would be a new approach to reveal spin-orbit-misaligned merging BH-NS systems, which are likely missed by the current gravitational-wave searching strategy principally targeting aligned systems.

\end{abstract} 


\section{Introduction}

The observation of the gravitational-wave (GW) source GW170817 \citep{abbott17a} by the Advanced LIGO \citep{ligo15} and the Advanced Virgo \citep{acernese15} GW detectors opened the multi-message astronomy era. Besides, its electromagnetic (EM) counterparts, including a short gamma-ray burst (sGRB), GRB 170817A \citep{abbott17b,goldstein17} with X-ray, optical and radio afterglows \citep{mooley18a,lyman18,troja17} and a kilonova, AT2017gfo \citep[e.g.,][]{abbott17c,arcavi17,coulter17,drout17,kasliwal17}, validated the hypothesis that binary neutron star (BNS) merger as an origin of the sGRB and an r-process production site. 

Similarly, black hole-neutron star (BH-NS) mergers should also be expected to produce EM counterparts \citep[e.g., kilonova and sGRB;][]{kyutoku13,barbieri19,li21}, as a result of the ejection of a non-negligible fraction of material when the NS is torn apart by the BH's tidal forces. The properties of the EM counterparts are sensitive to the mass and opacity of the ejected material, which depends on the spin of the BHs \citep{bauswein14,kyutoku15}, the equation of state (EoS) of NSs, and the magnetic field \citep{paschalidis15,kiuchi15}. However, the first-ever confidently detected two BH-NS merger events GW200105 and GW200115 \citep{abbott21} have no detected EM emission, despite the implementation of many post-GW-detection EM observation campaigns \citep[e.g.,][]{antier20,paterson21}. 

Many previous works tried to explain the absence of EM counterparts for those detected BH-NS merger GW events. Based on five GW events that are BH-NS merger candidates, \cite{zhu22} analyzed their mass and spin distributions, and found that most of such events should be plunging events (i.e., no tidal disruption-induced ejection of material) owing to their negligible spins. \cite{biscoveanu23} used four GW BH-NS events detected with false alarm rate $\leq1$ yr$^{-1}$ to constrain the mass and spin distributions and multi-messenger prospects of these systems. They found that BHs in BH-NSs are both less massive and more slowly spinning than those in binary black hole (BBH) mergers and no statistical preference between the two is found. Furthermore, they found that fewer than 14$\%$ of BH-NS mergers detectable in GWs will have an EM counterpart. \cite{chattopadhyay22} showed that the properties of a compact merger system critically depend on the formation channel of the system, and argued that a significant dominance of non-spinning BHs which were born before its companion NS in the merging BH-NS population explains the rare occurrence of EM counterparts.

Currently, the principal targets of GW searches are those compact binary objects that possess alignment between the spins of the binary components and the orbital angular momentum \citep{hooper12,harry14,usman16,messick17,aubin21}. This strategy leads to observational biases in the observed distribution of sources, whereas sources with non-negligible spin-orbit misalignment or unequal-mass ratio may be missed. Previous studies have examined the effect of the strategy on BBH searches \citep{harry16,harry18,bustillo17,chandra22}. Furthermore, \cite{dhurkunde22} investigated the bias in the observed BH-NS population. Specifically, simulating a population of BH-NS mergers whose spin distribution is isotropic in orientation and uniform in magnitude, they found that $\sim25\%$ of sources with mass-ratio $q>6$ and up to $\sim60\%$ of highly precessing sources ($\chi_{\rm p}>0.5$) might be missed. As a result, some BH-NS systems with high BH spin and spin-orbit misalignment, which potentially have EM counterparts, will be missed. This potentially explains the discrepancy that, currently, there are no detectable EM counterparts to the BH-NS merger GW events.

Despite that some BH-NS mergers with significant spin-orbit misalignment may have missed the GW window due to the biased searching strategy, they potentially have EM counterparts if the NS has been disrupted by a fast-spinning BH, and thus can be identified in the EM window alone. Here we propose that the observations of sGRBs resulting from the BH-NS mergers with high BH spin and significant spin-orbit misalignment serve as a measure for studying those GW-missed merger systems. Specifically, the misalignment would cause the resulting thick disk \citep[e.g.,][]{foucart11} precess in a solid-body manner around the remnant BH \citep{nelson00,fragile05,fragile07} due to the general relativistic Lense-Thirring (LT) torques \citep{lense18}. When the neutrino cooling in the disk becomes unimportant, the subsequently viscous-driven winds \citep[e.g.,][]{metzger08,metzger09,siegel18,de21,hayashi22} from the precessing disk will deflect the jet. The deflection causes the variation of the angle between the line of sight (LOS) and the jet, which would result in the occurrence of a quasi-periodic oscillation (QPO) in the sGRB light curve. This feature, unique to BH-NS mergers \citep{stone13}, might have already revealed itself in a few observed cases of GRBs, and it carries the spin-obit orientation information of the merging system. 

Previously, jet precession has already been considered in the context of sGRBs. Early works \citep[e.g.,][]{reynoso06,lei07,stone13} considered that the accretion disk formed after a merger behaves as a thick disk and precesses as a solid-body rotator due to the Lense-Thirring effect, and predicted the modulation to the light curve of a precessing jet. In the thin disk case, \cite{liu10} considered that the inner region of an inclined disk would experience a Bardeen-Petterson warp (the normal of the inner disk is along with the BH spin) and precess around the total angular momentum vector. Recently, jet precession is used to explain GRB 220408B ($T_{\rm 90,\,15-350 kev}\sim30$ s), which is observed to have a three-episode feature in its light curve \citep{zhang23}.

What sets the direction of the jet in GRBs is still ambiguous. \cite{liu10} considered that the jet originates from the neutrino-antineutrino annihilation mechanism \citep{meszaros92} and aligns with the angular momentum vector of the disk as the jet launching depends on the disk properties. \cite{reynoso06} and \cite{lei07} considered the jet launched from the Blandford-Znajek (BZ) mechanism \citep{blandford77} and also perpendicular to the disk plane as the magnetic field is anchored in the disk. Alternatively, \cite{stone13} consider the alignment of the jet with the BH spin axis, which is supported by the observation following the tidal disruption of a star by a supermassive BH \citep[Swift J164449.3$+$573451,][]{stone12}. The reason for this could be that the energy driving the Poynting-flux dominated jet comes from the spin of the BH \citep{beckwith08}. Thus, in this paper we consider a jet resulting from the BZ mechanism and having a BH-spin-aligned direction.

Here, for the first time we consider the effect of wind on the jet in BH-NS mergers, which is distinct from previous models. Due to the dominance of the angular momentum of the BH over that of the disk, the jet precesses around the spin axis of the BH after having been deflected by the wind from the precessing disk. The resulting QPO in the sGRB light curve would appear late, whose timing depends on the starting time of the wind ejection.

Note that our model is not for generally all sGRBs, but is unique to sGRBs from misaligned BH-NS mergers. For BNS mergers, in order to have a misaligned disk, which is vital to form a precessing disk around the central remnant BH and the resulting QPO in the sGRB, it requires that the NS should have a spin period shorter than 1 ms \citep{stone13}. However, the fastest spinning recycled NS is expected to have a period around of 4 ms \citep{willems08}. Therefore, the QPO proposed in our model is unlikely to appear in BNS mergers. 

This paper is structured as follows. In Section 2, we describe the model for the QPO appearing in the sGRB light curve of the misaligned BH-NS mergers in more detail, which is divided into six sub-sections. The prospect of observing such a late QPO in previous and future sGRBs is presented in Section 3. A case study for GRB 130310A with the appearance of the QPO is presented in Section 4. In Section 5, we summarize the results and discuss the implications.

\section{The Model}

\begin{figure}[t]
\begin{center}
\includegraphics[width=6.8cm, angle=0]{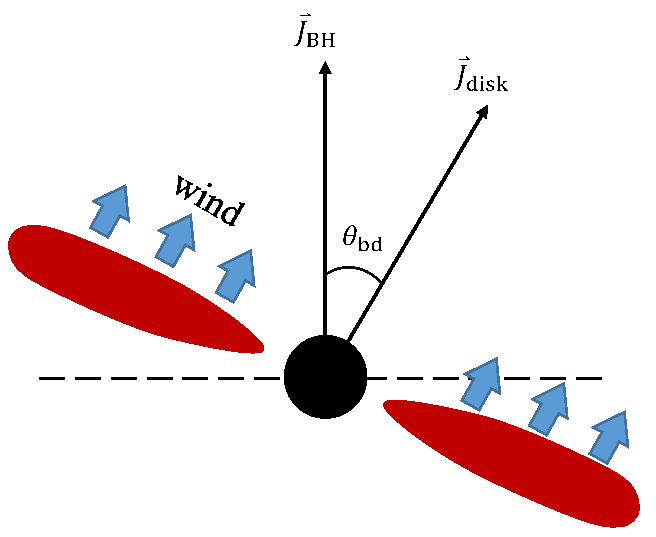}

\caption{Schematic illustration for the configuration of the BH-disk system (edge-on view) right after a BH-NS merger. Initially, a jet aligns with BH spin, $\bold{J}_{\rm BH}$, but the ejection of the wind (blue) from the disk (red) deflects and collimates the jet, causing the jet to align with the momentum of the disk, $\bold{J}_{\rm disk}$. $\theta_{\rm bd}$ represents the angle between $\bold{J}_{\rm BH}$ and $\bold{J}_{\rm disk}$, and is also the angle between $\bold{J}_{\rm BH}$ and the deflected jet.}\label{fig:bh_disk}
\end{center}
\end{figure}

\begin{figure}[t]
\begin{center}
\includegraphics[width=6.8cm, angle=0]{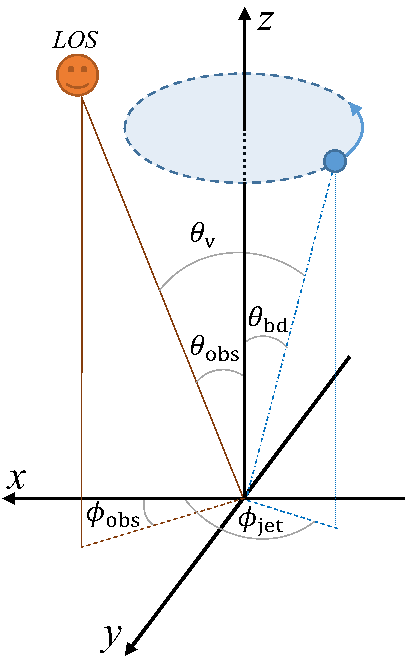}

\caption{Schematic illustration of the geometry of the model. The blue dashed-dotted line represents the axis of the deflected jet, which is inclined to the Z axis (equivalent to the spin axis of the BH) with an angle roughly equal to $\theta_{\rm bd}$. $\theta_{\rm obs}$ denotes the angle between the LOS and the spin axis of the BH. $\phi_{\rm obs}$ ($\phi_{\rm jet}$) denotes the angle between the projection of the LOS (axis of the deflected jet) and the arbitrary X axis. $\theta_{\rm v}$ represents the angle between the LOS and the jet axis, which varies with time due to the periodic change of $\phi_{\rm jet}$.}\label{fig:model}
\end{center}
\end{figure}

We establish a model to construct the sGRB light curve with the appearance of QPO. Initially, the GRB jet aligns with the BH spin vector, $\bold{J}_{\rm BH}$, while the disk precesses as a solid body ($\S$2.1). Subsequently, the wind ejection starts ($\S$2.2), and the wind deflects and collimates the jet ($\S$2.3), forcing the jet to align with the momentum of the disk, $\bold{J}_{\rm disk}$, as schematically illustrated in Fig.\ref{fig:bh_disk}. The jet axis rotates around the BH spin axis, resulting in the change of the viewing angle between the LOS and the jet axis ($\theta_{\rm v}$, Fig.\ref{fig:model}). The change of the viewing angle would cause the appearance of QPO in the late-time (at which the jet precesses owing to the wind) sGRB light curve, due to the dependence of the observed energies on the viewing angle ($\S$2.5).

The period and the occurring time of the QPO are estimated in $\S$2.1 and $\S$2.2, respectively. In $\S$2.3, we demonstrate that the viscous-driven wind from the precessing disk is powerful enough to deflect the jet. The jet structure and emission are described in $\S$2.4. In $\S$2.5, we describe in detail the method for modeling the QPO in the sGRB light curve. The numerical results are given in $\S$2.6.

\subsection{Precession of the accretion disk}

For a misaligned BH-NS merger, the newly formed accretion disk is initially tilted with respect to the spin axis of the BH remnant \citep{foucart11}. A test particle in a tilted orbit around a spinning BH would undergo the Lense-Thirring effect \citep{lense18}, resulting in the precession of the particle's orbital plane around the spin axis of the BH. With units $G=c=1$ ($G$ and $c$ are the gravitational constant and the speed of light, respectively), the angular speed of the precession is 
\begin{equation}
\bold{\Omega_{\rm LT}}= 2\bold{J_{\rm BH}}/R^3, 
\end{equation}
where $J_{\rm BH}=a_{\rm BH}M_{\rm BH}^2$ is the BH angular momentum, $a_{\rm BH}$ is the dimensionless BH spin parameter, $M_{\rm BH}$ is the mass of the BH, and $R$ is the orbit radius. Due to the strong dependence of $\Omega_{\rm LT}$ on $R$, the inner region of the tilted disk would experience a differential precession, hence inducing a warping in the accretion disk. 

The warping disturbance propagates outward in the disk in one of two manners, i.e., diffusive and wave-like. When the disk is sufficiently thin or highly viscous, i.e., $H/R<\alpha$, where $H$ and $\alpha$ represent the disk height and the dimensionless Shakura-Sunyaev viscosity coefficient, respectively, the propagation of warps will be diffusive \citep{papaloizou83}. Otherwise (i.e., a thick disk or $H/R>\alpha$), the warps would propagate in a wave-like manner, resulting in the redistributing of the torques throughout the disk \citep{papaloizou95,nelson99}.

It is expected that the propagation of the warp would stop at a radius where the local precession timescale ($\tau_{\rm LT}=\Omega_{\rm LT}^{-1}$) starts to exceed the local dynamical timescale (e.g., sound-crossing timescale), the latter of which depends on the propagation manner, i.e., diffusive or wave-like \citep{bardeen75,kumar85}. The stopping of the warp propagation results in the formation of the characteristic Bardeen-Petterson configuration in which the inner region of the disk aligns with the BH equatorial plane while the outer region keeps its original orientation \citep{bardeen75,papaloizou83,nelson00}. 

The transitioning radius from the inner aligned region to the outer inclined region depends on the thickness of the disk \citep{nelson00}. Particularly, \cite{nelson00} found that the thickest disk (with a mid-plane Mach number of 5) does not produce any discernible warped structure, i.e., no inner aligned region. Also, \cite{fragile05} and \cite{fragile07} found that the disk would undergo near rigid-body precession after a short initial period of differential precession, whenever the dynamical timescale is shorter than the precession timescale for the bulk of the disk. When this condition is met, the disk material is strongly coupled by the pressure waves, which causes the disk itself to act as a single entity in response to the torque of the BH and hence to precess as a solid body \citep{fragile07}.

Accretion disks formed in BH-NS mergers are always thick ($H/R>0.2$), almost independent of $R$, as reported in the full general relativity simulations \citep{foucart11}. Besides, the radius of the disk's peak surface density is $\sim50$ km \citep{foucart11}, indicating the compactness of the disk at an early time, which allows the above-mentioned condition to be fulfilled. Actually, \cite{foucart11} found that the precession angles in different radii tend to be equal (see their Fig. 13), which indicates that the disk finally undergoes near solid-body precession.
 
In the following, we estimate the disk precession period $T_{\rm prec}$, assuming the disk formed from a misaligned BH-NS merger has evolved quickly into the near solid-body precession state. In the Newtonian limit, $T_{\rm prec}=2\pi \sin\theta_{\rm bd}(J_d/\tau_d)$, where $J_d$ is the total angular momentum of the disk, $\theta_{\rm bd}$ is the angle between the angular momentum of the disk and the BH spin (Fig.\ref{fig:bh_disk}), and $\tau_d$ is the Lense-Thirring torque acting upon the disk \citep{liu02,fragile07,stone13,shen14}.

The total angular momentum of the disk can be expressed as \citep{shen14} 
\begin{equation}
{J_{\rm{d}}} = 2\pi {M_{\rm BH}^{{1 \mathord{\left/
 {\vphantom {1 2}} \right.
 \kern-\nulldelimiterspace} 2}}}\int_{{R_{\rm{i}}}}^{{R_{\rm{o}}}} {{{R}^{{3 \mathord{\left/
 {\vphantom {3 2}} \right.
 \kern-\nulldelimiterspace} 2}}}\Sigma \left( {R} \right)dR}
\end{equation}
where $\Sigma(R)$ represents the surface density at radius $R$, $R_{\rm{i}}$ and $R_{\rm{o}}$ are the radii of the inner and outer edges, respectively. The equivalent torque acting on the disk can be given as \citep{shen14}
\begin{equation}
\tau _{\rm{d}} = 2\pi J_{\rm BH}{M_{\rm BH}^{{1 \mathord{\left/
 {\vphantom {1 2}} \right.
 \kern-\nulldelimiterspace} 2}}}\sin \theta_{\rm bd} \int_{{R_{\rm{i}}}}^{{R_{\rm{o}}}} {{{R}^{{{ - 3} \mathord{\left/
 {\vphantom {{ - 3} 2}} \right.
 \kern-\nulldelimiterspace} 2}}}\Sigma \left( {R} \right)dR}. 
\end{equation}

Combining Eqs.(2) and (3), one arrives at
\begin{equation}
{T_{{\rm{prec}}}} = \frac{\pi }{J_{\rm BH}}\frac{{\int_{{R_{\rm{i}}}}^{{R_{\rm{o}}}} {{{R}^{{3 \mathord{\left/
 {\vphantom {3 2}} \right.
 \kern-\nulldelimiterspace} 2}}}\Sigma \left( {R} \right)dR} }}{{ \int_{{R_{\rm{i}}}}^{{R_{\rm{o}}}} {{{R}^{{{ - 3} \mathord{\left/
 {\vphantom {{ - 3} 2}} \right.
 \kern-\nulldelimiterspace} 2}}}\Sigma \left( {R} \right)dR} }}.
\end{equation}
Note that the units $G=c=1$ in Eqs.(1)-(4). 

Assuming a power-law surface density profile $\Sigma(R)=\Sigma_{\rm i} ({R}/{R_{\rm i}})^{-\zeta}$, one obtains 
\begin{equation}
{T_{{\rm{prec}}}} = \frac{{8\pi G{M_{{\rm{BH}}}}}}{{{a_{{\rm{BH}}}}{c^3}}}\left( {\frac{{2\zeta  + 1}}{{2\zeta - 5}}} \right)\frac{{r_{\rm{o}}^{{5 \mathord{\left/
 {\vphantom {5 2}} \right.
 \kern-\nulldelimiterspace} 2} - \zeta } - r_{\rm{i}}^{{5 \mathord{\left/
 {\vphantom {5 2}} \right.
 \kern-\nulldelimiterspace} 2} - \zeta }}}{{r_{\rm{o}}^{ - {1 \mathord{\left/
 {\vphantom {1 2}} \right.
 \kern-\nulldelimiterspace} 2} - \zeta } - r_{\rm{i}}^{{{ - 1} \mathord{\left/
 {\vphantom {{ - 1} 2}} \right.
 \kern-\nulldelimiterspace} 2} - \zeta }}},
\end{equation}
where $r_{\rm{o}}=R_{\rm o}/R_{\rm S}$, $r_{\rm{i}}=R_{\rm i}/R_{\rm S}$, $R_{\rm{S}}=2GM_{\rm BH}/c^2$ is the Schwarzschild radius of the BH. According to the simulations of misaligned BH-NS mergers performed by \cite{foucart11}, $\zeta$ roughly lies in the range of $0.5-1.5$ (see their Fig. 2). As given in Eq. (5), the precession period is sensitive to the size and surface density profile of the disk, i.e., $r_{\rm{o}}$, $r_{\rm{i}}$ and $\zeta$, and the spin and mass of the BH, but not to the mass of the disk. For the above mentioned typical range of $\zeta$, $T_{\rm prec}$ is approximated as $\propto r_{\rm o}^{5/2-\zeta}r_{\rm i}^{1/2+\zeta}$.

\begin{figure}[t]
\begin{center}
\includegraphics[width=8.7cm, angle=0]{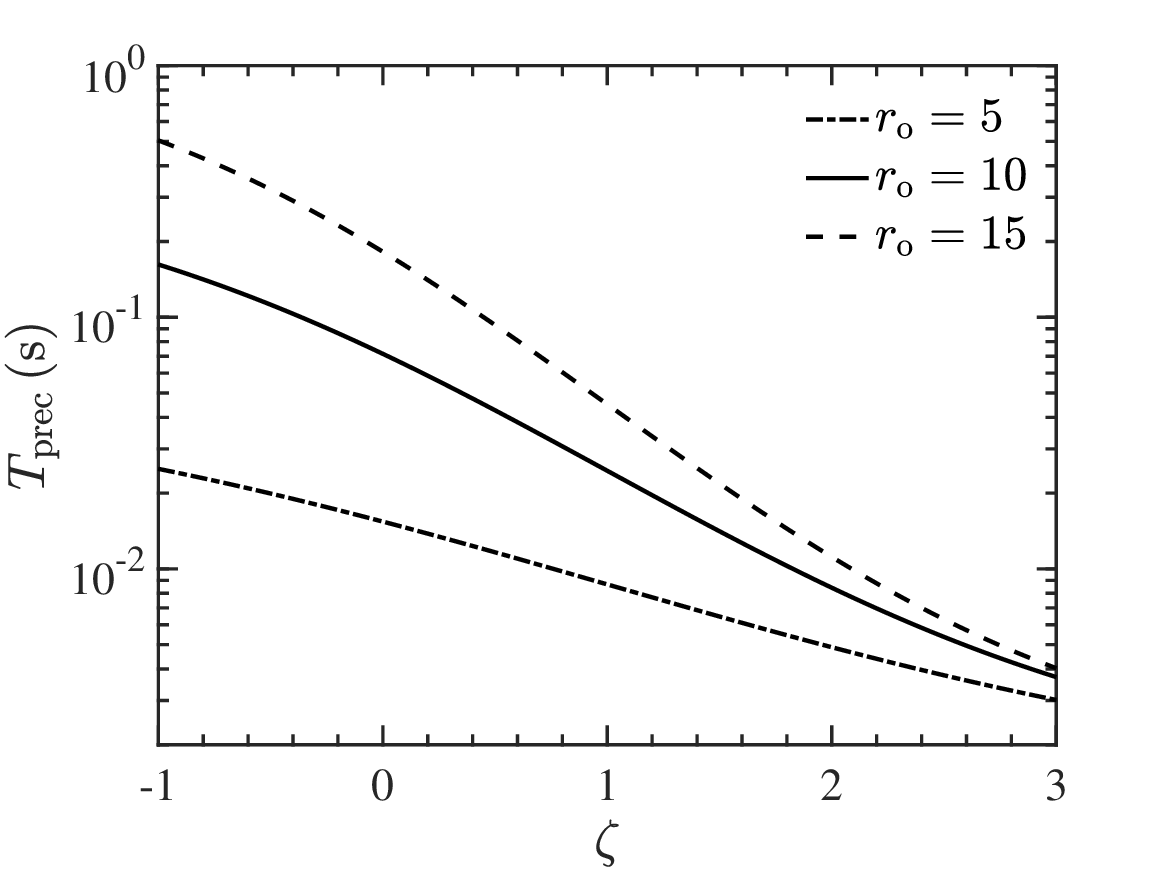}
\caption{The precession period of the accretion disk varying with $\zeta$. The outer radii for dashed, solid, dotted-dashed lines are 15$R_{\rm S}$, 10$R_{\rm S}$, 5$R_{\rm S}$, respectively. The mass and spin of the BH are set to 5$M_{\odot}$ and 0.8, respectively.}\label{fig:precession}
\end{center}
\end{figure}

Fig. \ref{fig:precession} shows the dependence of $T_{\rm prec}$ on the surface density profile for different outer radii (each line corresponds to a given outer boundary), with $M_{\rm BH}=5M_{\odot}$ and $a_{\rm BH}=0.8$. As the disk accretes, its surface density drops while the density profile mostly keeps unchanged \citep{foucart11}. As shown in Fig. \ref{fig:precession}, for the outer radii from $5R_{\rm S}$ to $15R_{\rm S}$ and the typical range of $\zeta$ ($0.5-1.5$), $T_{\rm prec}$ is estimated to locate in the range of $0.01-0.1$ s.

It should be noted that the disk also exerts the LT torque on the BH, resulting in the precession of the BH around the total angular momentum of the BH-disk system. The precession speeds of the BH and the disk, $\Omega_{\rm BH}$ and $\Omega_{\rm d}$, follow the relation as in $\Omega_{\rm BH}J_{\rm BH}=\Omega_{\rm d}J_{\rm d}$ \citep{tsokaros22}. According to Eq.(2), with $M_{\rm BH}=5M_{\odot}$, $a_{\rm BH}=0.8$, $M_{\rm disk}=0.1M_{\odot}$ and $r_{\rm o}=15$, one can roughly estimate the ratio $\Omega_{\rm BH} / \Omega_{\rm d}$ $=J_{\rm d} / J_{\rm BH}$ $\approx 0.1$. Therefore, the precession of the BH should be much slower than that of the disk. Besides, the BH spin would precess around the total angular momentum at a smaller angle compared with the disk angular momentum \citep{stone13}. Therefore, it is difficult to observe the spin-induced precession of the jet unless the jet is deflected by the wind as proposed in our work.
 
\subsection{(Delayed) ejection of disk wind}

A hot, thick and hyperaccreting disk, as the case in sGRBs, may be cooled via neutrino emission when the accretion rate $\dot{M}_{\rm acc}$ is sufficiently high \citep{popham99,narayan01,beloborodov03,chen07}. As the sufficiently dense mid-plane of the hyperaccreting disk allows electrons to reach degeneracy, weak interactions are ignited, which result in the thermal neutrino emission and affect the thermaldynamics and composition of the disk \citep{chen07,metzger09,siegel18}. Such disks are classified as neutrino dominated accretion flows (NDAFs).

The threshold of accretion rate for igniting the weak interactions, $\dot{M}_{\rm ign}$, has been studied numerically \citep{chen07} and analytically \citep{de21}. The results show that $\dot{M}_{\rm ign}$ depends on the mass and spin of the BH as well as the viscosity coefficient of the disk, which can be expressed as \citep{chen07, de21}
\begin{equation}
{\dot M_{{\rm{ign}}}} \approx {\dot {\cal M}_{{\rm{ign}}}}\left( {{M_{{\rm{BH}}}},{a_{{\rm{BH}}}}} \right){\alpha ^{{5 \mathord{\left/
 {\vphantom {5 3}} \right.
 \kern-\nulldelimiterspace} 3}}}.
\end{equation}
\cite{de21} analytically derive the scaling relationship $\dot{\cal M}_{\rm ign}\propto M_{\rm BH}^{4/3}$. For a BH with spin of 0.8, the ignition threshold can be estimated as \citep{de21}
\begin{equation}
{\dot M_{{\rm{ign}}}} \approx 2 \times {10^{ - 3}}{M_ \odot }{{\rm{s}}^{ - 1}}{\left( {\frac{{{M_{{\rm{BH}}}}}}{{3{M_ \odot }}}} \right)^{{4 \mathord{\left/
 {\vphantom {4 3}} \right.
 \kern-\nulldelimiterspace} 3}}}{\left( {\frac{\alpha }{{0.02}}} \right)^{{5 \mathord{\left/
 {\vphantom {5 3}} \right.
 \kern-\nulldelimiterspace} 3}}}.
\end{equation}

\begin{figure}[t]
\begin{center}
\includegraphics[width=8.7cm, angle=0]{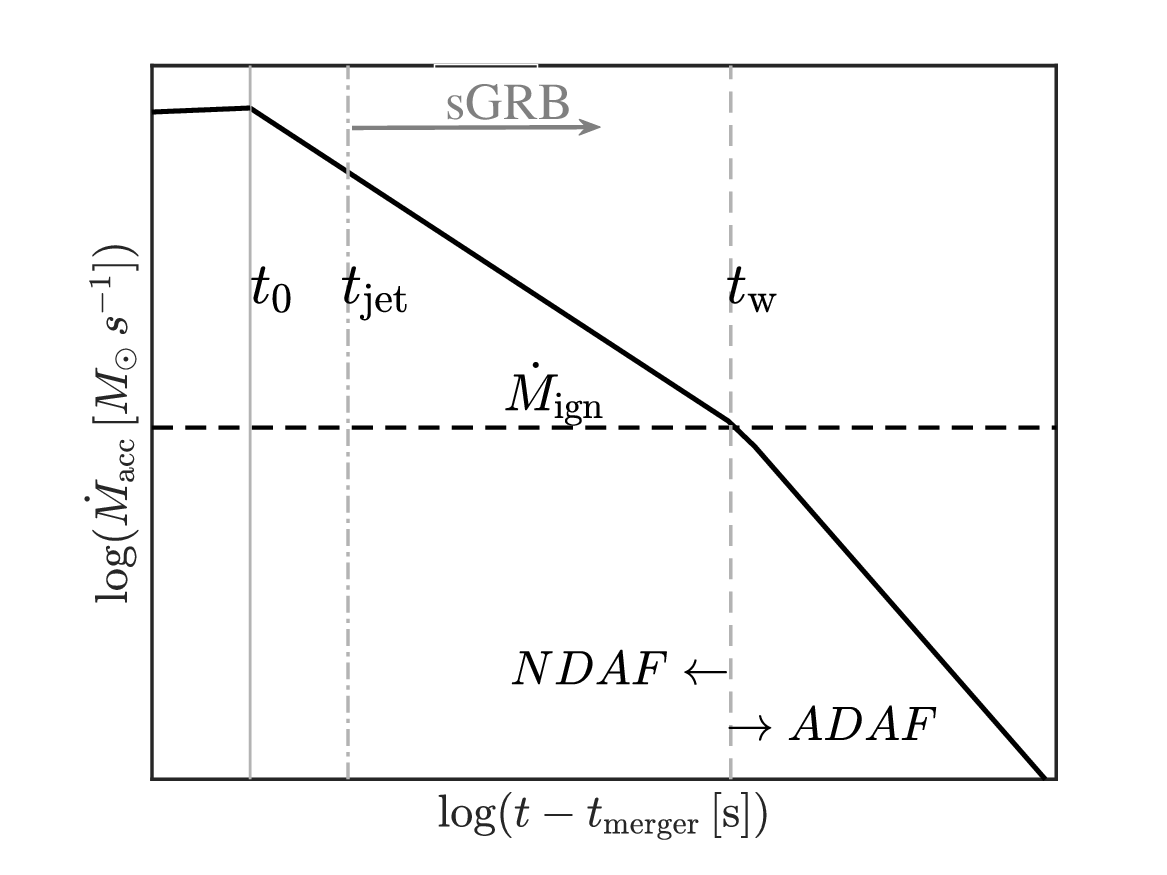}
\caption{Schematic diagram for the evolution of the accretion rate (the black solid line) after a compact binary merger. After reaching the maximum accretion rate at $t=t_0$ (the gray solid line), the accretion rate starts to decrease with time as $\dot{M}_{\rm acc} \propto t^{-\eta}$. Initially, as the accretion rate is higher than $\dot{M}_{\rm ign}$ (the black dashed line), the disk stays in the neutrino-cooled state. Once the accretion rate evolves to be lower than $\dot{M}_{\rm ign}$, the disk enters into the ADAF state and the viscous-driven wind is ejected. The intersection between the black solid and black dashed lines shows the starting time for the ejection of disk wind, which is denoted as $t_{\rm w}$ and marked by the gray dashed line. Due to the ejection of the wind, the accretion rate for the ADAF state drops more sharply compared with that for the NDAF state. The gray dashed-dotted line ($t_{\rm jet}$) denotes the time for launching the relativistic jet.}\label{fig:accretion_rate}
\end{center}
\end{figure}

During the initial hyperaccretion onto the central BH, the disk undergoes viscous expansion due to angular momentum transport, resulting in the decrease of $\dot{M}_{\rm acc}(t)$. Once $\dot{M}_{\rm acc}(t)$ drops below $\dot{M}_{\rm ign}$, both the weak interactions and the neutrino cooling are no longer dominant. As a result, the thermal energy resulted from viscous heating in the accretion disk could not be radiated away efficiently, which significantly influences the subsequent evolution of the disk. As the post-merger disk has transitioned to an advective dominated accretion flow (ADAF), a powerful viscous-driven wind is ejected \citep[e.g.,][]{metzger08,metzger09,siegel18,de21,hayashi22}.

Some simulations for the evolution of the neutrino-cooled accretion disks from compact binary mergers found that $\dot{M}_{\rm acc}(t)$ quickly reaches a peak value at 10's of ms after the merger and then decreases monotonically with time approximately as $\propto t^{-\eta}$ \citep[e.g.,][]{de21,hayashi22}. Therefore, one can express the accretion rate as
\begin{equation}
\dot{M}_{\rm acc}=\frac{M_0}{t_0}(\frac{t}{t_0})^{-\eta}
\end{equation}
where $M_0$ is the initial mass of disk, and $t_0$ denotes the beginning of the decrease of the accretion rate. 

Fig. \ref{fig:accretion_rate} schematically shows the evolution of the accretion rate (the black solid line) for a disk formed after a BH-NS merger. The black dashed line denotes the ignition threshold for neutrino cooling. Thus the intersection between the black solid and black dashed lines roughly marks the starting time for the ejection of the powerful viscous-driven wind 
\begin{equation}
t_{\rm w}=(\frac{M_{\rm 0}}{t_{\rm 0}\dot{M}_{\rm ign}})^{\frac{1}{\eta}}t_{\rm 0}.
\end{equation}
Simulations for the neutrino-cooled accretion disk show $t_{\rm w}\sim0.1$ s \citep[e.g.,][]{siegel18,de21,hayashi22}. After $t_{\rm w}$, when $\dot{M}_{\rm acc}(t)$ evolves to below $\dot{M}_{\rm ign}$, as the disk transitions from NDAF to ADAF state, $\dot{M}_{\rm acc}(t)$ drops more steeply than before due to the ejection of wind \citep{metzger09,siegel18,de21,hayashi22}. 

The launching time $t_{\rm jet}$ of the jet roughly corresponds to the instrument trigger time of the sGRB, which depends on the jet launching mechanism. In the case of BH-NS mergers, $t_{\rm jet}$ is $\sim10$ ms for a jet launched via the BZ mechanism \citep[e.g.,][]{zhang19,sun22}. Simulations suggest $t_{\rm jet}>t_0$ \citep[e.g.,][]{hayashi22}. The gray dashed-dotted line in Fig. \ref{fig:accretion_rate} denotes $t_{\rm jet}$. Note that the exact value of $t_{\rm jet}$ is not critical in our model as long as $t_{\rm jet}<t_{\rm w}$.

\subsection{Jet deflection by the wind}

Even in the ADAF state, the accretion disk still satisfies the condition (i.e., $H/R>\alpha$) to precess as a rigid body for the following two reasons. Firstly, the majority of the disk still possesses a large ratio of $H/R$ ($>0.3$) after the transition from neutrino-cooled state to advective state \citep{metzger09,kumar15}. Secondly, the viscosity parameter $\alpha$ roughly keeps constant and small, $\sim 0.01-0.1$ for different simulation runs \citep{de21}.

As the disk precesses with a period of $T_{\rm prec}$, so does the symmetry axis, or the total momentum vector, of the wind. The latter may deflect the jet and force it to precess with the same period.

To demonstrate the capability of the wind that can induce jet precession, here, we simply compare their momentum fluxes. We estimate the isotropic-equivalent momentum per unit time (momentum flux) of the jet and the wind, respectively, as
\begin{equation}
{{\dot P}_{{\rm{jet}}}} = \frac{L_{\rm jet, iso}}{c} \approx 3 \times {10^{39}} \left( {\frac{{{L_{{\rm{jet}},{\rm{iso}}}}}}{{{{10}^{50}}{{{\rm{erg}}} \mathord{\left/
 {\vphantom {{{\rm{erg}}} {\rm{s}}}} \right.
 \kern-\nulldelimiterspace} {\rm{s}}}}}} \right){{{\rm{erg}}} \mathord{\left/
 {\vphantom {{{\rm{erg}}} {{\rm{cm}}}}} \right.
 \kern-\nulldelimiterspace} {{\rm{cm}}}}
\end{equation}

\begin{equation}
{{\dot P}_{\rm{w}}} \approx 6 \times {10^{41}}\left( {\frac{{{{\dot M}_{\rm{w}}}}}{{0.1{{{M_ \odot }} \mathord{\left/
 {\vphantom {{{M_ \odot }} {\rm{s}}}} \right.
 \kern-\nulldelimiterspace} {\rm{s}}}}}} \right)\left( {\frac{{{{\rm{v}}_{\rm{w}}}}}{{0.1c}}} \right){{{\rm{erg}}} \mathord{\left/
 {\vphantom {{{\rm{erg}}} {{\rm{cm}}}}} \right.
 \kern-\nulldelimiterspace} {{\rm{cm}}}}
\end{equation}
where $L_{\rm jet,iso}$ and ${\dot M}_{\rm w}$ denote the isotropic-equivalent luminosity of the jet and the ejection rate of the wind, respectively, $\rm{v}_{\rm w}$ is the wind velocity. The value of ${\dot M}_{\rm w}$ is referred to the simulations performed for neutrino-cooled disks formed after BH-NS mergers \citep{hayashi22}. With the typical values of $L_{\rm jet,iso}$ and ${\dot M}_{\rm w}$, the momentum flux of the wind is about two orders of magnitude larger than that of the jet. As a result, the viscous-driven wind potentially deflects the jet to almost align with the wind direction (i.e., the normal of the disk), causing the appearance of QPO in the GRB light curve.

The above treatment (comparing momentum fluxes) is certainly over-simplified. A more careful way of investigating the capability of deflection would be to run magneto-hydrodynamic simulation of the interaction, such as done by \cite{ohsuga23} in a similar context.

\subsection{Jet angular structure and emission}

Previous studies of GRB 170817A, one of the electromagnetic counterparts to the BNS merger event GW170817, have revealed the presence of a structured jet \citep{mooley18a,mooley18b,ghirlanda19}. The angular structure of a jet is characterized by the jet energy per unit solid angle, $dE/d\Omega = \epsilon$, and the Lorentz factor, $\Gamma$, both of which vary with the angle $\theta$ from the jet axis (see below). For a structured jet, the observed energy from the resulting GRB under different viewing angles $\theta_{\rm v}$, i.e., the angle between the LOS and the jet axis, would be different \citep{lipunov01}. Note that in the following all quantities are measured in the central engine's reference frame. The cosmological redshift effect of the GRB host is neglected.

The energy and Lorentz factor profiles of a jet structure were often described by power-law \citep{meszaros98,rossi02,zhang02} and Gaussian models \citep{zhang02,kumar03,zhang04}. As the actual jet structure of sGRB is still a controversial topic \citep[e.g.,][]{nakar18,takahashi21}, we consider a Gaussian structure in this work, for simplicity. Therefore, the energy and the Lorentz factor profiles are described as \citep{zhang02,kumar03,salafia15}
\begin{equation}
\begin{split}
& \epsilon(\theta)=\epsilon_c e^{-(\theta /\theta_c)^2}\\
& \Gamma(\theta)=1+(\Gamma_c-1) e^{-(\theta /\theta_c)^2}
\end{split}
\end{equation}
where $\theta_c$ is the typical angular scale measured from the jet axis. Most of the jet energy is contained within $\theta_c$, hence $\theta_c$ can be roughly considered as the half opening angle of the jet core. 

\begin{figure}[t]
\begin{center}
\includegraphics[width=8.7cm, angle=0]{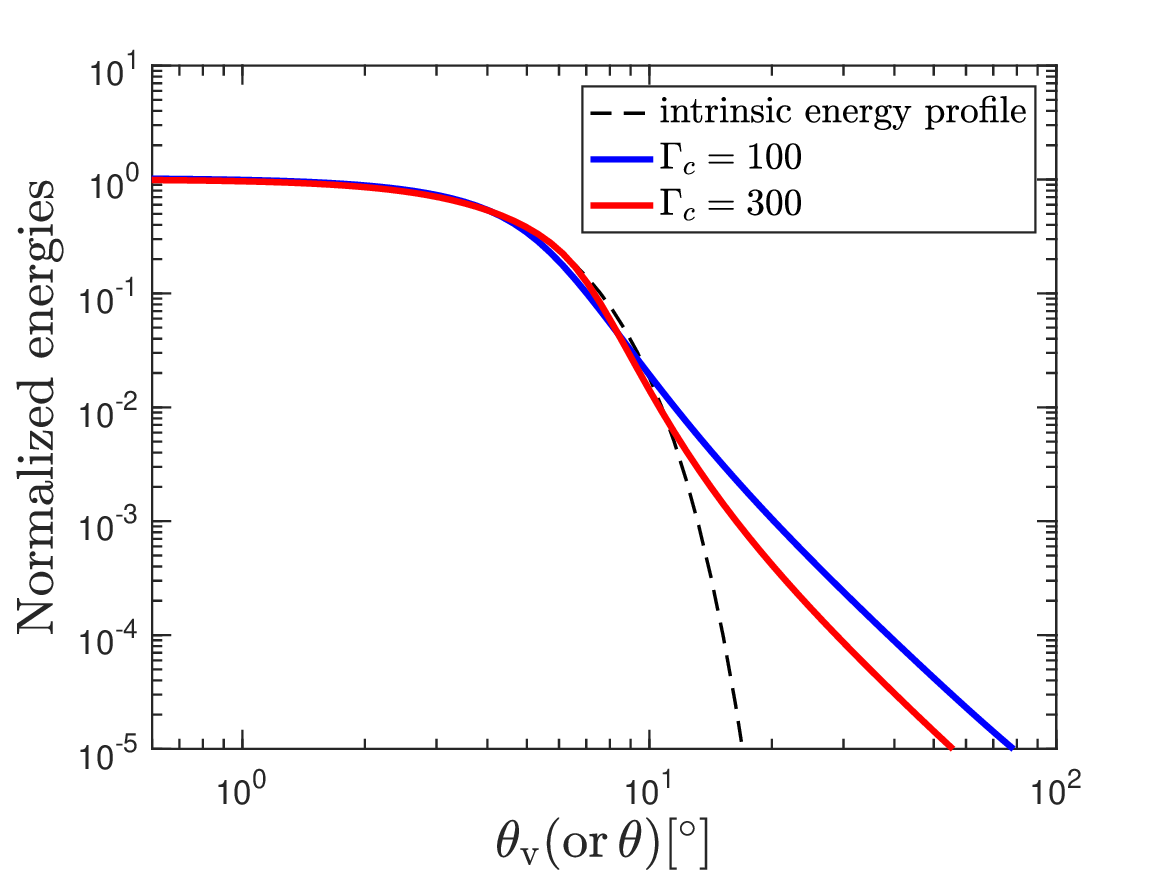}
\caption{The observed isotropic equivalent energies normalized to their corresponding maximum values as a function of the viewing angle $\theta_{\rm v}$ for the Gaussian jet model (red and blue). The blue and red lines correspond to $\Gamma_c$ =100 and 300, respectively. The half opening angle of the jet core $\theta_c$ is set to 5$^\circ$. The normalized intrinsic energy profile for the Gaussian jet, i.e., $e^{-(\theta/\theta_{\rm c})^2}$, is shown as the dashed line for comparison.}\label{fig:SJ}
\end{center}
\end{figure}

We adopt a method introduced by \cite{salafia15} to derive the dependence of the observed energies on the viewing angle. The observed isotropic equivalent energy from a structured jet for an observer at a given viewing angle $\theta_{\rm v}$ can be derived as \citep{salafia15}
\begin{equation}
{E_{{\rm{iso}}}}\left( {{\theta _{\rm{v}}}} \right) = \int\limits_{\phi = 0}^{2\pi } {\int\limits_{\theta = 0}^{{\pi \mathord{\left/
{\vphantom {\pi 2}} \right.
\kern-\nulldelimiterspace} 2}} {\frac{{{\delta ^3}\left( {\theta ,\phi ,{\theta _{\rm{v}}}} \right)}}{{\Gamma \left( \theta \right)}}\epsilon \left( \theta \right){\rm{sin}}\theta {\rm{d}}\theta {\rm{d}}\phi } }
\end{equation}
and the Doppler factor is \citep{salafia15}
\begin{equation}
\delta(\theta,\phi,\theta_{\rm v})=\frac{1}{\Gamma(\theta)[1-\beta(\theta){\rm cos}\,\mu(\theta,\phi,\theta_{\rm v})]},
\end{equation}
where $\phi$ is the azimuth angle, $\beta(\theta)$ represents the velocity that corresponds to $\Gamma(\theta)$, and $\mu$ denotes the angle between the velocity of the emitting material and the observer's LOS, which can be derived as \citep{salafia15}
\begin{equation}
{\rm cos}\,\mu(\theta,\phi,\theta_{\rm v})= {\rm cos}\,\theta\,{\rm cos}\,\theta_{\rm v}+{\rm sin}\,\theta\,{\rm sin}\,\phi\, {\rm sin}\,\theta_{\rm v}.
\end{equation}

Combining Eqs. (12)--(15), one can compute the observed isotropic equivalent energy from a Gaussian jet as a function of the viewing angle. Fig. \ref{fig:SJ} shows the examples for a Gaussian jet with $\theta_{\rm c}=5^{\circ}$. The blue and red lines correspond to $\Gamma_c=$ 100 and 300, respectively. As shown in Fig. \ref{fig:SJ}, when $\theta_{\rm v}<\theta_{\rm c}$ (within the jet core), the observed energies roughly keep constant, approximately the same as the plateau appearing in the intrinsic energy profile (the dashed line). When $\theta_{\rm v}>\theta_{\rm c}$, the observed energies decay with $\theta_{\rm v}$ and the larger the Lorentz factor is, the steeper the decay is.

Note that our above treatment adopts a time-independent jet angular structure. In reality, the strong oblique shocks induced in the jet-wind interaction would probably alter the jet structure in a time-dependent manner \citep[e.g.,][]{mizuta13}. Nevertheless, by simply considering the forced precession of the jet axis, our model illustrates the emergence of QPOs in GRB light curves, exemplified by the case of GRB 130310A (see below). A more sophisticated treatment considering the jet-structure change might fully comprehend the intricacies of the observed data.

\subsection{GRB light curve and periodic modulation}

In this work, we consider the jet resulted from BZ mechanism \citep{blandford77} and having a BH-spin-aligned direction. However, once the jet is deflected by the wind from the precessing disk, i.e., $t>t_{\rm w}$, its axis would begin to rotate around the spin axis of the central BH with a precession rate $\Omega_{\rm prec}$, resulting in the variation of the viewing angle of the jet for a given observer. Therefore, the observed GRB light curve would display certain periodic features resulted from the precession of the disk. 

As shown in Fig. \ref{fig:model}, the angle between the LOS and the spin axis of the BH is denoted as $\theta_{\rm obs}$. The angle between the projection of the LOS to the X axis is denoted as $\phi_{\rm obs}$. Due to the wind deflection, the axis of the deflected jet (blue dashed-dotted line) is inclined off the spin axis of the BH with an angle roughly equal to $\theta_{\rm bd}$. The angle between the projection of the jet axis and the X axis is denoted as $\phi_{\rm jet}$. $\theta_{\rm obs}$ and $\phi_{\rm obs}$ are fixed for a given sGRB. The projection of the jet axis in the XY plane is rotating around the original point with a rate equal to the precessing rate of the disk, thus the varying angle $\phi_{\rm jet}$ is derived as 
\begin{equation}
{\phi _{{\rm{jet}}}} = \left\{ {\begin{array}{*{20}{c}}
{{\phi _{\rm{jet,0}}}}&{t < {t_{{\rm{w}}}}}\\
{{\phi _{\rm{jet,0}}} + \frac{{2\pi t}}{{{T_{{\rm{prec}}}}}}}&{t \ge {t_{{\rm{w}}}}}.
\end{array}} \right.
\end{equation}

The viewing angle of the jet axis, $\theta_{\rm v}$ (shown in Fig. \ref{fig:model}), can be computed as
\begin{equation}
\begin{array}{l}
\cos \theta_{\rm v} = \cos {\theta _{{\rm{obs}}}}\cos {\theta _{{\rm{bd}}}} + \\
\quad \quad \;\;\,\sin {\theta _{{\rm{obs}}}}\sin {\theta _{{\rm{bd}}}}\cos \left( {{\phi _{{\rm{jet}}}} - {\phi _{{\rm{obs}}}}} \right),
\end{array}
\end{equation}
which varies with time as well. It should be noted that the value of $\theta_{\rm v}$ is equal to $\theta_{\rm obs}$ when $t<t_{\rm w}$, due to the alignment of the initial BZ jet with the spin axis of the BH. 

The received flux from a given sGRB can be described as
\begin{equation} \label{eq:F}
{\cal{F}}(t)=I(t)\,E_{\rm iso}(\theta_{\rm v})/(4\pi \epsilon_{\rm c})
\end{equation}
where $I(t)$ is the intrinsic (i.e., observed always on-axis, or $\theta_{\rm v}\equiv0$) energy intensity of the jet. $E_{\rm iso}(\theta_{\rm v})/(4\pi \epsilon_{\rm c})$ accounts for the dependence of the received flux on $\theta_{\rm v}$, which is normalized as shown in Fig.\ref{fig:SJ}.

For demonstrative purpose, we use a piecewise power law function to describe time dependence of the intrinsic energy intensity as in
\begin{equation}
I\left( t \right) = \left\{ {\begin{array}{*{20}{c}}
{\propto{t^{\alpha_{1}}}}&{t < {t_{{\rm{p}}}}}\\
{\propto{t^{-\alpha_{2}}}}&{t \ge {t_{{\rm{p}}}}}
\end{array}} \right.
\end{equation}
where $\alpha_{1}>0$, and $\alpha_{2}>0$ are constants, and $t_{\rm p}$ denotes the time when the intensity reaches its maximum. In real cases of GRB light curves, more complex temporal profiles might be needed.

\begin{figure}[t]
\begin{center}
\includegraphics[width=8.7cm, angle=0]{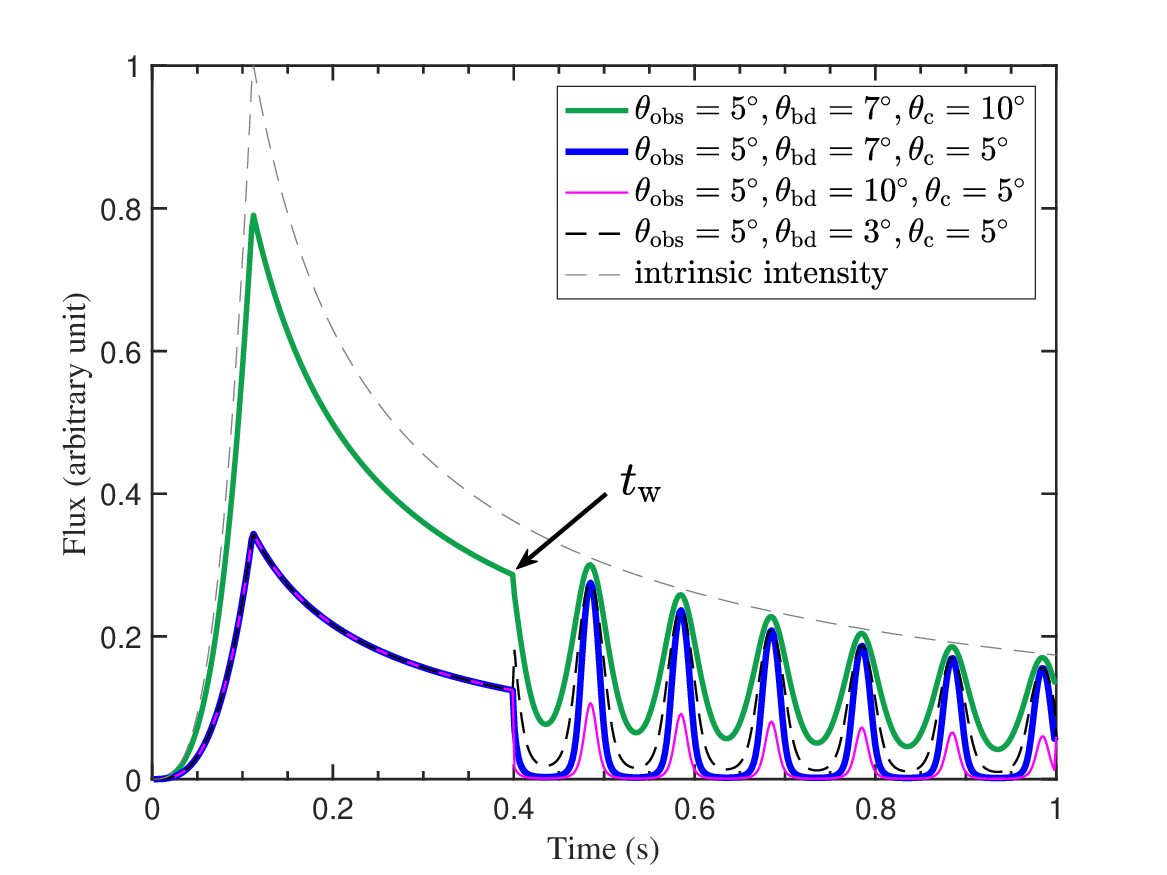}

\caption{The predicted sGRB light curve with the appearance of the QPO. We compute 4 different sets of $\theta_{\rm obs}$, $\theta_{\rm bd}$, and $\theta_c$ (i.e., the half opening angle of the jet core), whose exact values are shown in the legend. $\phi_{\rm obs}$ and $\phi_{\rm jet,0}$ are set to 0 and $0.3\pi$, respectively. The precession period is set to 0.1 s. The starting time for the wind ejection, i.e., the appearing time for the QPO, is set to 0.4 s. We adopt a Gaussian jet with $\Gamma_c=100$. The intrinsic intensity (Eq. 19) is normalized to 1 (grey dashed line).}\label{fig:model_qpo}
\end{center}
\end{figure}

The time-varying relativistic Doppler effect not only affects the radiation energy (as represented in Eq. 13), but also alters the temporal structure of the energy intensity from the precessing jet. However, the latter effect is not considered in Eq. (18). We anticipate that this effect would increase the amplitude of the QPO profile and cause a slight asymmetry of the profile. For more strict calculation, we refer the reader to previous works \citep[e.g.,][]{liu10,zhang23}

\subsection{Numerical results}

With the model constructed above, Fig. \ref{fig:model_qpo} shows four examples of sGRB light curves with four sets of values of $\theta_{\rm obs}$, $\theta_{\rm bd}$, and $\theta_c$. The grey dashed line denotes the intrinsic intensity $I(t)$ (Eq. 19), whose peak is normalized to 1. We set $\phi_{\rm obs}=0$, $\phi_{\rm jet,0}=0.3\pi$, the starting time for the wind ejection $t_{\rm w}$ (equivalent to the QPO appearing time) to be 0.4 s, and the precession period to be 0.1 s. 

As shown in Fig. \ref{fig:model_qpo}, after the wind ejection, i.e., $t>t_{\rm w}$, the light curve starts to display periodic features, which are mainly determined by the variation of $\theta_{\rm v}$. The peak of each oscillation decreases with time, due to the decay of the sGRB light curve. All peaks (and the troughs) are appearing at the same time due to the same precession rate and $\phi_{\rm jet,0}$. The peak (or trough) is determined by the minimum (or maximum) value of $\theta_{\rm v}$, which depends on $\theta_{\rm obs}$, $\theta_{\rm bd}$ and $\phi_{\rm jet}$. 

It is important to note that the wind ejection time $t_{\rm w}$ depends on the initial mass of the disk and the evolution of the accretion rate. If the initial accretion rate is very close to the ignition rate, the resulting QPO would appear earlier. Thus the observer may detect the QPO with a longer duration and higher intensity.

Fig. \ref{fig:model_qpo} also shows the parameter dependences of the QPO amplitude (relative to the main peak of the GRB light curve). For the three cases with $\theta_{\rm c}=5^{\circ}$ (i.e., the blue, magenta, and black-dashed lines), their light curves are the same before the occurrence of the disk wind (i.e., $t<t_{\rm w}$). After the wind deflection of the jet (i.e., $t>t_{\rm w}$), their corresponding QPO amplitudes decrease with increasing viewing angle offsets (i.e., $|\theta_{\rm bd}-\theta_{\rm obs}|$). For the case of $\theta_{\rm c}=10^{\circ}$ (i.e., the green line), its QPO amplitude is much stronger than that of the other three cases, because a wider jet core results in an observed intensity that is closer to the peak intensity, $\epsilon_{\rm c}$ (note that $\epsilon_{\rm c}$ keeps constant in all cases while we change the value of $\theta_{\rm c}$).

Note that the disk's precession period is expected to grow instead of keeping constant, because the outer boundary of the disk still increases after the wind ejection (c.f., Eq. (16) \& Fig. \ref{fig:precession}). As a result, the period of the oscillation shown in Fig. \ref{fig:model_qpo} for each case should increase with time. Besides, the inclination of the angular momentum of the disk with respect to the spin axis of the BH ($\theta_{\rm bd}$) tends to decrease with time \citep{foucart11}. Thus at a sufficiently late time, the inclination may become so small that the precession of the disk terminates, and hence the QPO in the GRB light curve would disappear.

\begin{figure}[t]
\begin{center}
\includegraphics[width=8.7cm, angle=0]{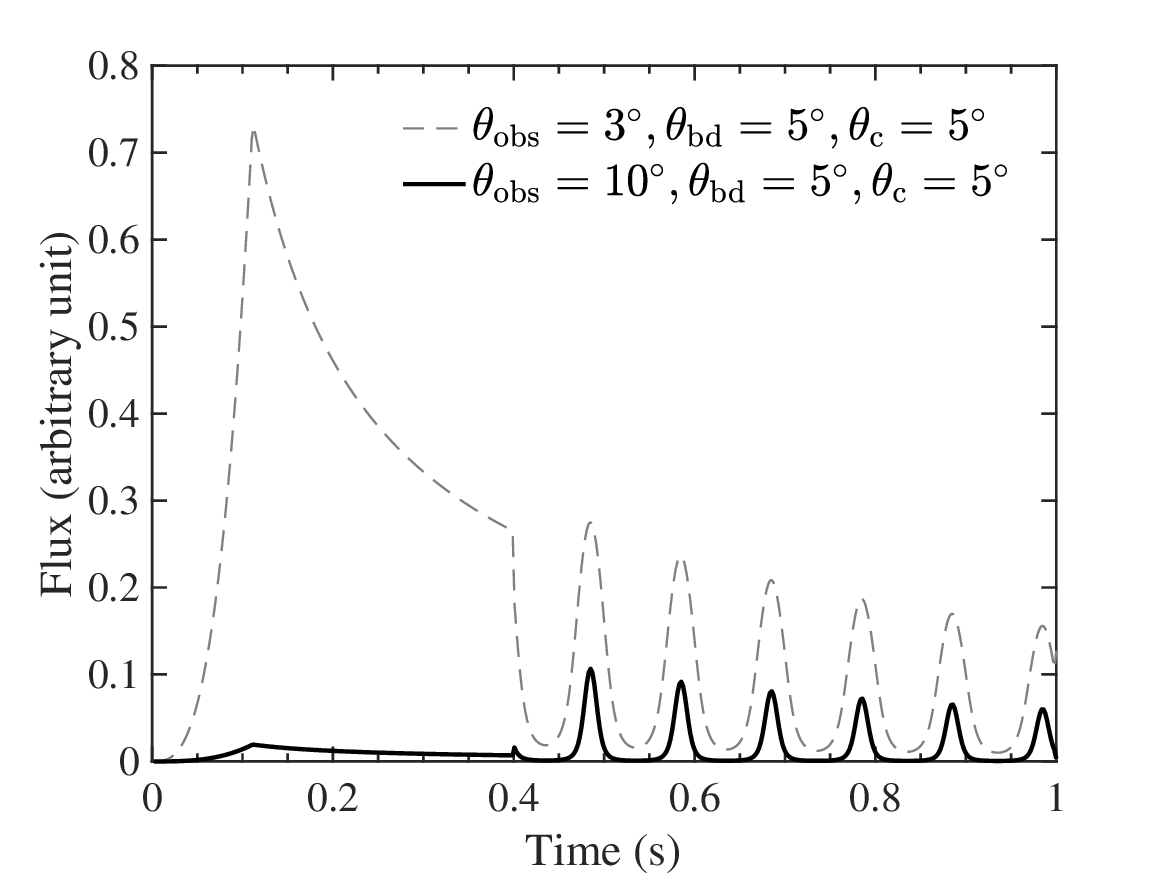}

\caption{The two typical sGRB light curves with the appearance of the QPO. The black solid line represents the scenario that the early main burst of an off-axis sGRB could be invisible but followed by an observable QPO due to the precessing jet. The light curve of an on-axis sGRB with an observable main burst (the grey dashed line), which is the same as the green line in Fig. \ref{fig:model_qpo}, is shown for comparison.
}\label{fig:main_burst_off_axis}
\end{center}
\end{figure}

\section{An orphan QPO -- Precession of an initially off-axis jet}

The results shown in Fig. \ref{fig:model_qpo} represent the typical light curve pattern of sGRBs. They are initially observed on-axis with significant main bursts followed by QPOs with lower intensity due to the wind-induced jet precession. 

There remains another possibility that some sGRBs are off-axis (i.e., with a significant misalignment between the LOS and the jet) and hence invisible initially. However, when the deflected jet starts to precess around the spin axis of the BH, the observer can have the chance to periodically see it, which behaves as a quasi-periodic bursting source for a short duration. 

Taking a relatively large value of $\theta_{\rm obs}$, the black solid line in Fig. \ref{fig:main_burst_off_axis} shows the case that the early main burst is invisible but the QPO is occurring. This is obviously different from the grey dashed line with an observable main burst, which is the same as the green line in Fig. \ref{fig:model_qpo}. The difference is determined by whether $\theta_{\rm obs}$ is larger than $\theta_{\rm c}$.

\section{Case Study}

\begin{figure*}[ht]
\begin{center}
\subfigure{
\includegraphics[width=8.7cm]{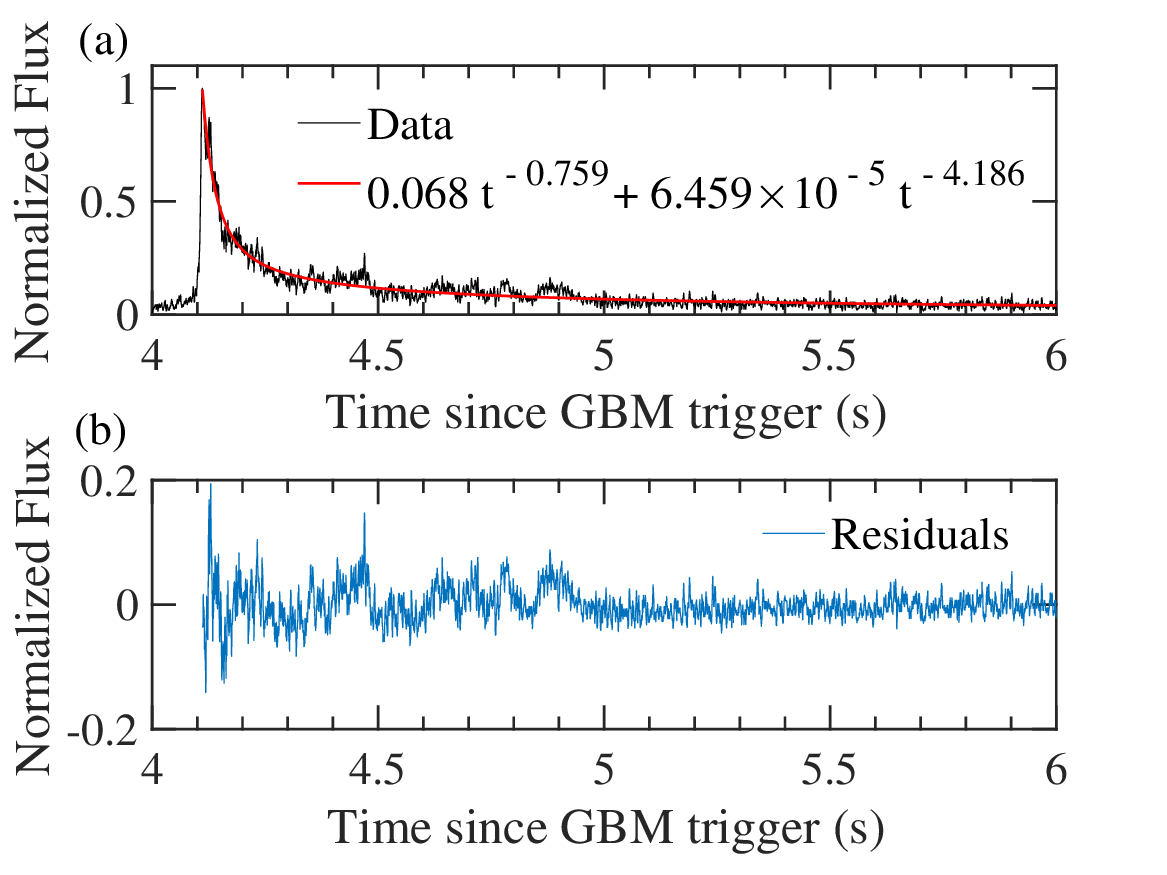}}
\subfigure{
\includegraphics[width=8.7cm]{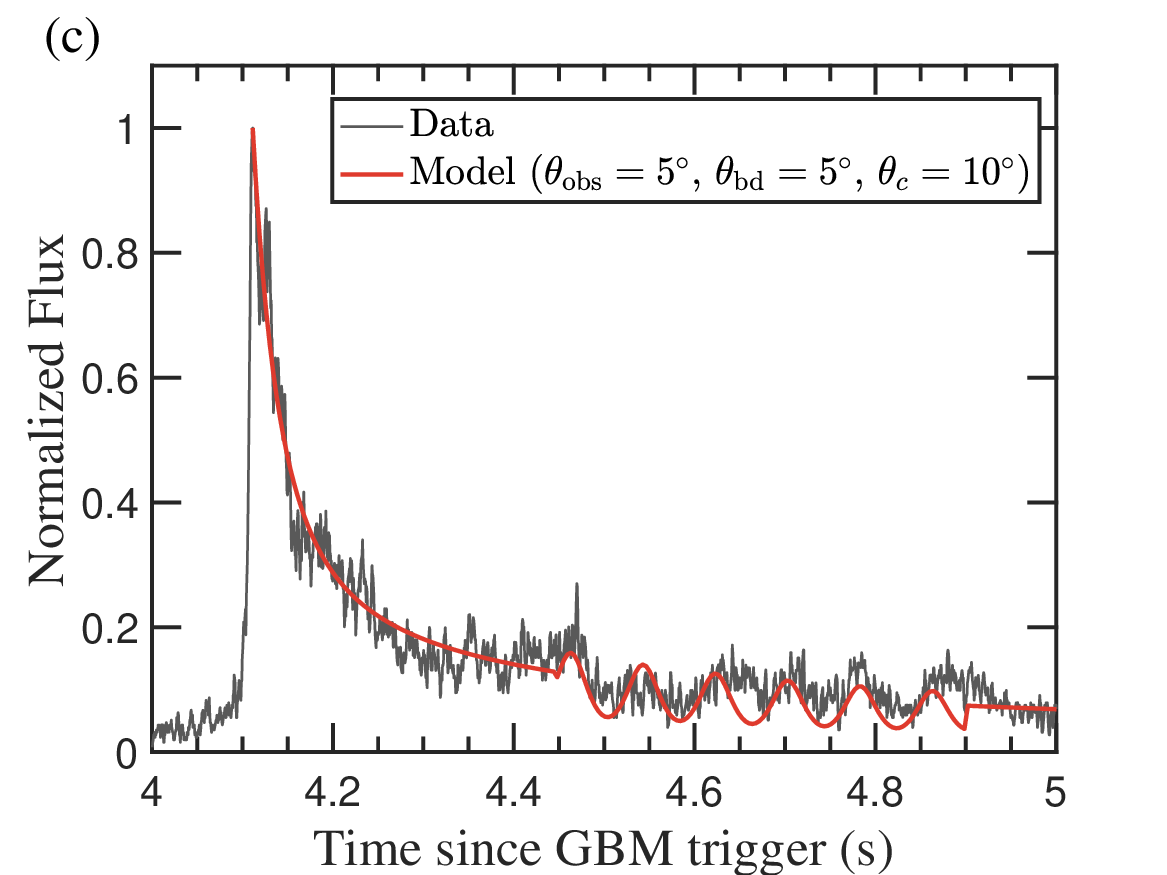}}
\end{center}
\caption{Light curve and QPO modeling of GRB 130310A. \textbf{Panel (a):} To clearly show the QPO feature, the raw light curve data of GRB 130310A is firstly smoothed through a 5-point moving average (the black line). Then we use the sum of two power laws (the red line), to de-trend the decay phase of the light curve. This gives a better fit to the decaying trend than either a single power law or an exponential. \textbf{Panel (b):} The residuals of the de-trending, which show a QPO signature during $T_0+(4.4\sim4.9)\,{\rm s}$. \textbf{Panel (c):} Lastly, we apply out jet precession model (Eq. \ref{eq:F}, the red line) to fit the overall decaying light curve, where the temporal profile of the jet intrinsic radiative intensity adopts the two-power law function in panel (a). The precession period is 80 ms. }\label{fig:fitting}
\end{figure*}

The model constructed here demonstrates the appearance of a delayed QPO in the prompt emission of sGRB with different light curve patterns, which is unique to the BH-NS merger. The most direct observational signature of jet precession in our model is the prediction of quasi-period around $0.01-0.1$ s. Searching for such QPO is helpful to identify the merging binary and derive the spin-orbit orientation information of the merging system. 

Claim of periodicity in sGRBs has been rare, but recently a weak periodic oscillation is found in GRB 130310A, with a period of $\sim80$ millisecond \citep{zhang22}. This burst contains two major emission episodes, i.e., a precursor at the trigger time $T_0$ and a main burst at $T_0+3.8\,{\rm s}$ \citep{zhang22}. The $T_{\rm 90}$ of the precursor and the main burst is 1.3 s and 2.9 s, respectively. 

To show this feature of late oscillations, we use a 5-point moving average to smooth the raw data of \cite{zhang22} so as to suppress the high frequency noise, as shown in Fig. \ref{fig:fitting}a and \ref{fig:fitting}c (i.e., the black line). The main burst displayed a sharp peak and was followed by a relatively soft tail. Then we use the sum of two power law functions to de-trend the data within $4.1-6$ s (the red line in Fig. \ref{fig:fitting}a; the resulting R-square, 0.94), as it is superior to either single power law or exponent function. The de-trended light curve shows an oscillation during $T_0+(4.4\sim4.9)\,{\rm s}$ (Fig. \ref{fig:fitting}b), which is consistent with \cite{zhang22}. 

We propose that the GRB 130310A arise from a BH-NS merger, as an alternative to the explanation by \cite{zhang22} that it is a giant flare from a newly born magnetar. Here, we do the modeling of the late QPO feature. We apply our precession model (i.e., Eq. \ref{eq:F}) to best fit the light curve with QPO occurring between 4.4 and 4.9 s since the trigger, at which the periodic emission is most significant. The total modeled flux with $T_{\rm prec}=80$ ms \citep{zhang22}, $\theta_{\rm obs}=5^{\circ}$, $\theta_{\rm bd}=5^{\circ}$ and $\theta_{\rm c}=10^{\circ}$ matches well with the observational data (Fig. \ref{fig:fitting}c).

There remains another set of parameters (i.e., $\theta_{\rm obs}=3^{\circ}$, $\theta_{\rm bd}=2^{\circ}$ and $\theta_{\rm c}=5^{\circ}$) that can also match the data well, which is almost the same as the red line in Fig. \ref{fig:fitting}c (not shown for clarity). The distinct two sets of parameters display similar pattern, indicating the degeneracy of the parameters. Though the appearance of the main burst suggests $\theta_{\rm obs}$ should not be much larger than $\theta_{\rm c}$, the lack of constraint on $\theta_{\rm c}$ causes a difficulty in distinguishing the two sets.

The major advantage of our proposed scenario is that the emergence of the viscous-driven wind as discussed in $\S2.2$ is the natural cause of the delay between the appearance of the QPO and the peak of the main burst, $\sim0.4$ s as identified in the GRB light curve (Fig. \ref{fig:fitting}). Besides, the disappearance of QPO after 4.9 s can be explained by the final alignment of the disk and the BH. However, the other scenario (i.e., a giant flare of a newly born magnetar) may be a little bit difficult to explain why the rotation-caused $\sim80$ ms period only appears between 4.4 and 4.9 s.

The period of the QPO is around 80 ms, which locates in the estimated regions of the precession period for a disk with $\zeta \sim 0.5-1.5$ and $r_{\rm o} \sim 10-15$, as shown in Fig. \ref{fig:fitting}c. It is worth noting that the duration of each circle is not exactly equal to a fixed value, but increases with time. This is consistent with the prediction given at the end of $\S2.6$, i.e., the precession period of a disk resulted from a misaligned BH-NS merger would be enlarged due to the evolution of the disk with its outer boundary gradually increasing. Fixing the period as 80 ms during the whole emergence episode of the QPO causes that since the third circle, the model does not fit the data well.

As the redshift of GRB 130310A as well as its luminosity distance have not been measured yet \citep{minaev17,zhang22}, we take its total fluence of ${\cal F}_{\rm t}=4.12^{+0.32}_{-0.32}\times10^{-5}$ erg cm$^{-2}$ \citep{zhang22} and the typically observed isotropic equivalent energy of the prompt $\gamma$-ray emission of sGRBs with $E_{\rm \gamma,iso}\sim10^{49}-10^{52}$ erg \citep{berger14} to infer the probable luminosity distance ${D_{\rm{L}}} \simeq {\left( {\frac{{{E_{\gamma ,iso}}}}{{4\pi {{\cal F}_{\rm{t}}}}}} \right)^{1/2}}$ to be $0.046-1.47$ Gpc. Adopting a Hubble constant $H_{\rm 0}=68.8$ km s$^{-1}$ Mpc$^{-1}$ \citep{gray22}, the corresponding redshift is estimated to be $0.0108-0.3394$, which overlaps with that of the scenario for GRB 130310A being a MGF GRB, $0.0057-0.1187$ \citep{zhang22}, whereas it is also well within the range of sGRB redshift, $0.1-1.5$ \citep{berger14}. As a result, the scenario that GRB 130310A is a sGRB from a BH-NS merger cannot be ruled out.

The duration of the precursor ($T_{\rm pre}$) and the main burst ($T_{\rm GRB}$) are equal to 1.3 and 2.9 s based on $T_{\rm 90}$, respectively, and the waiting time of the main burst ($T_{\rm wt}$) is around 3.8 s. Current theoretical models of sGRB precursors \citep[e.g.,][]{hansen01,wang18,zhang22b} are difficult to explain its waiting time, as it is expected to shorter than that from BNS mergers (e.g., the waiting time of the GRB 170817A from a BNS merger is 1.7 s) due to the lack of the collapse of an NS to form a BH. However, the distribution of waiting time of BNS mergers is elusive due to the insignificant statistics (i.e., only one case, GRB 170817A, has been observed). Nevertheless, these times roughly satisfy the requirement for the sGRBs that $T_{\rm pre}\sim T_{\rm GRB}\sim T_{\rm wt}$ \citep{wang20}.

\section{Summary and Discussion}

Despite that some BH-NS mergers with significant spin-orbit misalignment may have missed the GW window due to the biased searching strategy, they potentially have EM counterparts if the NS has been disrupted by a fast-spinning BH, and thus can be identified in the EM window alone. 

Due to their naturally unequal component masses, the post-merger BH and the NS-debris disk would keep the initial misalignment between the components' angular momenta. Therefore, the disk would precess as the result of the Lense-Thirring effect. In this work, we propose that a jet with a spin-aligned direction resulted from the BZ mechanism is deflected and collimated by the wind from the precessing disk, which results in a QPO in the light curve of the resulting sGRB. Identifying this feature from sGRB light curves would serve as supporting evidence for a BH-NS merger.

We have constructed a model for the QPO appearing in the light curve of sGRB from BH-NS mergers. The predicted QPO is mainly dependent of the half opening angle $\theta_{\rm c}$ of the jet core, the inclined angle $\theta_{\rm bd}$ between the angular momenta of the BH and the disk, and the angle $\theta_{\rm obs}$ between the LOS and the spin axis of the BH. The main predictions of our model are:
\par
(1) The periods of the QPOs are in the range of $\sim10-100$ ms (c.f., Fig. \ref{fig:precession});
\par
(2) The typical light curve pattern has a prominent peak followed by a dimmer QPO. The delayed QPOs are the natural consequence of the wind ejection when the disk is evolving into the ADAFs state. 
\par
(3) If the jet is initially off-axis (i.e., $\theta_{\rm obs}>\theta_{\rm c}$) and later its precession sweeps across the line of sight, then one would see an orphan QPO without a main burst. 
\par
(4) A more significant QPO (compared with the corresponding main burst) can result from smaller $|\theta_{\rm bd}-\theta_{\rm obs}|$ and $\theta_{\rm c}$ (c.f., Fig. \ref{fig:model_qpo}). 

A few factors might cause the QPOs predicted here too weak to be identified. Firstly, the misalignment $\theta_{\rm bd}$ should not be too small, otherwise the QPO would become insignificant. Secondly, if the outer radius of the disk is too large, e.g., due to viscous spreading, the precession period may become longer than the duration of the prompt emission, then the delayed QPO would barely show up. Thirdly, the intrinsic variability of the jet luminosity could be mixed with the QPO feature and make it hard to distinguish the two.

We find GRB 130310A to be a probable candidate. Applying our jet precession model to it, we find a good match between the model and the observation. Both two sets of parameters (one is $\theta_{\rm obs}=5^{\circ}$, $\theta_{\rm bd}=5^{\circ}$ and $\theta_{\rm c}=10^{\circ}$, and the other is $\theta_{\rm obs}=3^{\circ}$, $\theta_{\rm bd}=2^{\circ}$ and $\theta_{\rm c}=5^{\circ}$) can match well with the observed data. Our model can explain GRB 130310A as all the constrained parameters are within the previously-constrained range, such as $2^{\circ}<\theta_{\rm c}<10^{\circ}$ \citep{kumar15}.

Recently, \cite{chirenti23} found kilohertz-QPOs identified in the short bursts GRB 910711 and GRB 931101B from archival Burst and Transient Source Experiment (BATSE) data. The periods of the QPOs in GRB 910711 and GRB 931101B ($\sim0.4$ ms) are much shorter than that predicted in our model which is unique to the misaligned BH-NS merger. Those kilohertz-QPOs are suggested to be produced by the fast oscillation of a proto-NS newly born from a BNS merger \citep{chirenti23}.


\section*{}
YL and RFS are supported by National Natural Science Foundation of China (No. 12073091) and China Manned Spaced Project (CMS-CSST-2021-B11). BBZ is supported by National Key Research and Development Programs of China (2018YFA0404204, 2022YFF0711404, 2022SKA0130102), National SKA Program of China (2022SKA0130100), National Natural Science Foundation of China (Nos. 11833003, U2038105, U1831135, 12121003), China Manned Space Project (CMS-CSST-2021-B11), Fundamental Research Funds for the Central Universities, Program for Innovative Talents and Entrepreneur in Jiangsu, and Chinese Academy of Sciences, No. XDB23040400.



\end{CJK*}

\begin{thebibliography}{99}

\bibitem[Abbott et al.(2017a)]{abbott17a} Abbott, B. P., Abbott, R., Abbott, T. D., et al., 2017a, PhRvL, 119, 161101

\bibitem[Abbott et al.(2017b)]{abbott17b} Abbott, B. P., Abbott, R., Abbott, T. D., et al., 2017b, ApJL, 848, L13

\bibitem[Abbott et al.(2017c)]{abbott17c} Abbott, B. P., Abbott, R., Abbott, T. D., et al., 2017c, ApJL, 848, L12

\bibitem[Abbott et al.(2021)]{abbott21} Abbott R., Abbott, R., Abbott, T. D., et al., 2021, Astrophys. J. Lett., 915, L5

\bibitem[Acerness et al. (2015)]{acernese15} Acernese, F., Agathos, M., Agatsuma, K., et al., 2015, CQGra, 32, 024001

\bibitem[Antier et al.(2020)]{antier20} Antier, S., Agayeva, S., Almualla, M., et al., 2020, MNRAS, 497, 5518

\bibitem[Tsokaros et al.(2022)]{tsokaros22} Tsokaros, A., Ruiz, M., Shapiro, S. L., Paschalidis, V., 2022, PRD, 106, 104010

\bibitem[Arcavi et al.(2017)]{arcavi17} Arcavi, I., Hosseinzadeh, G., Howell, D. A., et al., 2017, Natur, 551, 64

\bibitem[Aubin et al.(2021)]{aubin21} Aubin F., Brighenti F., Chierici R., et al., 2021, Class. Quantum Grav. 38 095004

\bibitem[Barbieri et al.(2019)]{barbieri19} Barbieri, C., Salafia, O. S., Perego, A., Colpi, M., \& Ghirlanda, G. 2019, A\&A,625, A152

\bibitem[Bardeen \& Petterson(1975)]{bardeen75} Bardeen J. M. \& Petterson J. A., 1975, Astrophys. J. Lett., 195, L65 .

\bibitem[Bauswein et al.(2014)]{bauswein14} Bauswein, A., Ardevol Pulpillo, R., Janka, H. T., \& Goriely, S. 2014, ApJL, 795, L9

\bibitem[Beckwith et al.(2008)]{beckwith08} Beckwith, K., Hawley, J. F., \& Krolik, J. H., 2008, ApJ, 678, 1180-1199 

\bibitem[Beloborodov(2003)]{beloborodov03} Beloborodov, A. M. 2003, ApJ, 588, 931

\bibitem[Berger(2014)]{berger14} Berger, E. 2014, ARA\&A, 52, 43

\bibitem[Blandford \& Znajek(1977)]{blandford77} Blandford, R. D., \& Znajek, R. L. 1977, MNRAS, 179, 433

\bibitem[Biscoveanu et al.(2023)]{biscoveanu23} Biscoveanu S., Landry P., Vitale S., 2023, MNRAS, 518, 4, 5298-5312

\bibitem[Bustillo et al.(2017)]{bustillo17} Bustillo J. C., Laguna P., \& Shoemaker D., 2017, Phys. Rev. D, 95, 104038

\bibitem[Chandra et al.(2022)]{chandra22} Chandra K., Bustillo J. C., Pai A., \& Harry I. W., 2022, Phys. Rev. D 106, 123003 

\bibitem[Chattopadhyay et al. (2022)]{chattopadhyay22} Chattopadhyay D., Stevenson S., Broekgaarden F., 2022, MNRAS, 513, 4, 5780

\bibitem[Chen \& Beloborodov(2007)]{chen07} Chen, W.-X., \& Beloborodov, A. M. 2007, ApJ, 657, 383

\bibitem[Chirenti et al.(2023)]{chirenti23} Chirenti, C., Dichiara, S., Lien, A., Miller, M. C., Preece, R., 2023, Nature, 613, 7943, 253-256

\bibitem[Coulter et al.(2017)]{coulter17} Coulter, D. A., Foley, R. J., Kilpatrick, C. D., et al., 2017, Sci, 358, 1556

\bibitem[De \& Siegel(2021)]{de21} De, S., Siegel, D. M., 2021, ApJ, 921, 1, 94

\bibitem[Dhurkunde \& Nitz(2022)]{dhurkunde22} Dhurkunde R., \& Nitz, A. H., 2022, Phys. Rev. D, 106, 10, 103035

\bibitem[Drout et al.(2017)]{drout17} Drout, M. R., Piro, A. L., Shappee, B. J., et al. 2017, Sci, 358, 1570

\bibitem[Foucart et al.(2011)]{foucart11} Foucart F., Duez M. D., Kidder L. E., and Teukolsky S. A., 2011, Phys. Rev. D, 83, 024005.

\bibitem[Fragile \& Anninos(2005)]{fragile05} Fragile, P. C., \& Anninos, P. 2005, ApJ, 623, 347

\bibitem[Fragile et al.(2007)]{fragile07} Fragile, P. C., Blaes, O. M., Anninos, P., \& Salmonson, J. D. 2007, ApJ, 668, 417

\bibitem[Ghirlanda et al.(2019)]{ghirlanda19} Ghirlanda G. et al., 2019, Science, 363, 968

\bibitem[Goldstein et al.(2017)]{goldstein17} Goldstein A., Veres P., Burns E., et al., 2017, ApJL, 848, L14

\bibitem[Gray et al.(2022)]{gray22} Gray, R., Messenger, C., \& Veitch, J., 2022, MNRAS, 512, 1, 1127-1140

\bibitem[Hansen \& Lyutikov(2001)]{hansen01} Hansen, B. M. S., \& Lyutikov, M. 2001, MNRAS, 322, 695

\bibitem[Harry et al.(2014)]{harry14} Harry I. W., Nitz A. H., Brown, D. A., 2014, Phys. Rev. D 89, 024010

\bibitem[Harry et al.(2016)]{harry16} Harry I., Privitera S., Boh\'{e} A., \& Buonanno A., 2016, Phys. Rev. D, 94, 024012

\bibitem[Harry et al.(2018)]{harry18} Harry I., Bustillo J. C., \& Nitz A., 2018, Phys. Rev. D, 97, 023004

\bibitem[Hayashi et al.(2022)]{hayashi22} Hayashi, K., Fujibayashi, S., Kiuchi, K., Kyutoku, K., Sekiguchi, Y., Shibata, M., 2022, Physical Review D, 106, 2, 023008

\bibitem[Hooper et al.(2012)]{hooper12} Hooper S., Chung S. K., Luan J., et al., 2012, Phys. Rev. D, 86, 024012

\bibitem[Kasliwal et al.(2017)]{kasliwal17} Kasliwal, M. M., Nakar, E., Singer, L. P., et al., 2017, Sci, 358, 1559

\bibitem[Kiuchi et al.(2015)]{kiuchi15} Kiuchi, K., Sekiguchi, Y., Kyutoku, K., et al., 2015, PhRvD, 92, 064034

\bibitem[Kumar \& Pringle(1985)]{kumar85} Kumar, S., \& Pringle, J. E., 1985, MNRAS, 213, 435

\bibitem[Kumar \& Granot(2003)]{kumar03} Kumar P., \& Granot J., 2003, ApJ, 591, 1075

\bibitem[Kumar \& Zhang(2015)]{kumar15} Kumar, P., \& Zhang, B., 2015, Physics Reports, 561, 1-109. 

\bibitem[Kyutoku et al.(2013)]{kyutoku13} Kyutoku, K., Ioka, K., \& Shibata, M. 2013, PhRvD, 88, 041503

\bibitem[Kyutoku et al.(2015)]{kyutoku15} Kyutoku, K., Ioka, K., Okawa, H., Shibata, M., \& Taniguchi, K. 2015, PhRvD, 92, 044028

\bibitem[Lense \& Thirring(1918)]{lense18} Lense, J., \& Thirring, H. 1918, Phys. Z., 19, 156

\bibitem[LIGO Scientific Collaboration et al.(2015)]{ligo15} LIGO Scientific Collaboration, Aasi, J., Abbott, B. P., et al., 2015, CQGra, 32, 074001

\bibitem[Lei et al.(2007)]{lei07} Lei W. H., Wang D. X., Gong B. P., \& Huang C. Y., 2007, A\&A, 468, 563-569

\bibitem[Li \& Shen(2021)]{li21} Li, Y., \& Shen, R.-F. 2021, ApJ, 911, 87

\bibitem[Lipunov et al.(2001)]{lipunov01} Lipunov V. M., Postnov K. A., Prokhorov M. E., 2001, Astron. Rep., 45, 236

\bibitem[Liu \& Melia(2002)]{liu02} Liu, S., \& Melia, F. 2002, ApJ, 573, L23

\bibitem[Liu et al.(2010)]{liu10} Liu T., Liang E.-W., Gu W.-M., Zhao X.-H., Dai Z.-G., \& Lu J.-F., 2010, A\&A. 516, A16

\bibitem[Lyman et al.(2018)]{lyman18} Lyman J. D., Lamb G. P., Levan A. J., et al., 2018, Nature Astronomy, 2, 751-754 

\bibitem[Messick et al.(2017)]{messick17} Messick C., et al., 2017, Phys. Rev. D 95, 042001

\bibitem[Metzger et al.(2008)]{metzger08} Metzger B. D., Piro A. L., Quataert E., 2008, MNRAS, 390, 781

\bibitem[Metzger et al.(2009)]{metzger09} Metzger, B. D., Piro, A. L., \& Quataert, E. 2009, MNRAS, 396, 304

\bibitem[M\'{e}sz\'{a}ros \& Rees(1992)]{meszaros92} M\'{e}sz\'{a}ros P., \& Rees M. J., 1992, MNRAS, 257, 29P

\bibitem[M\'{e}sz\'{a}ros et al.(1998)]{meszaros98} M\'{e}sz\'{a}ros P., Rees M. J., Wijers R. A. M. J., 1998, ApJ, 499, 301

\bibitem[Minaev \& Pozanenko(2017)]{minaev17} Minaev, P. Y., \& Pozanenko, A. S. 2017, AstL, 43, 1

\bibitem[Mizuta \& Ioka(2013)]{mizuta13} Mizuta, A., \& Ioka, K., 2013, ApJ, 777, 162

\bibitem[Mooley et al.(2018a)]{mooley18a} Mooley K. P. et al., 2018, Nature, 554, 207

\bibitem[Mooley et al.(2018b)]{mooley18b} Mooley K. P. et al., 2018b, Nature, 561, 355

\bibitem[Narayan et al.(2001)]{narayan01} Narayan, R., Piran, T., \& Kumar, P. 2001, ApJ, 557, 949

\bibitem[Nakar \& Piran (2018)]{nakar18} Nakar E., Piran T., 2018, MNRAS, 478, 407

\bibitem[Nelson \& Papaloizou (1999)]{nelson99} Nelson, R. P., \& Papaloizou, J. C. B. 1999, MNRAS, 309, 929

\bibitem[Nelson \& Papaloizou (2000)]{nelson00} Nelson, R. P., \& Papaloizou, J. C. B. 2000, MNRAS, 315, 570

\bibitem[Norris et al.(1996)]{norris96} Norris J. P., Nemiroff R. J., Bonnell J. T., et al. 1996. ApJ. 459:393

\bibitem[Ohsuga (2023)]{ohsuga23} Ohsuga, K., 2023, presentation at ``Astrophysical Black Holes: A Rapidly Moving Field'', University of Hong Kong, Hong Kong, China, 23-26 June 2023, https://astrobh.physics.hku.hk/event/3/contributions/13/

\bibitem[Paterson et al.(2021)]{paterson21} Paterson, K., Lundquist, M. J., Rastinejad, J. C., et al. 2021, ApJ, 912, 128

\bibitem[Papaloizou \& Lin(1995)]{papaloizou95} Papaloizou J. C. B., \& Lin D. N. C., 1995, ApJ, 438, 841

\bibitem[Papaloizou \& Pringle(1983)]{papaloizou83} Papaloizou J. C. B., Pringle J. E., 1983, MNRAS, 202, 1181

\bibitem[Paschalidis et al.(2015)]{paschalidis15} Paschalidis, V., Ruiz, M., \& Shapiro, S. L. 2015, ApJL, 806, L14

\bibitem[Popham et al. (1999)]{popham99} Popham R., Woosley S. E., \& Fryer C. L., 1999, ApJ, 518, 356

\bibitem[Reynoso et al.(2006)]{reynoso06} Reynoso M. M., Romero G. E., \& Sampayo O. A., 2006, A\&A, 454, 11-16

\bibitem[Rossi et al.(2002)]{rossi02} Rossi E., Lazzati D., Rees M. J., 2002, MNRAS, 332, 945

\bibitem[Salafia et al.(2015)]{salafia15} Salafia O. S., Ghisellini G., Pescalli A., Ghirlanda G., Nappo F., 2015, MNRAS, 450, 3549

\bibitem[Usman et al.(2016)]{usman16} Usman S. A., Nitz A. H., Harry I. W., 2016, Class. Quant. Grav., 33, 215004

\bibitem[Shen \& Matzner(2014)]{shen14} Shen, R.-F., \& Matzner, C. D., 2014, ApJ, 784, 2, 87

\bibitem[Siegel \& Metzger(2018)]{siegel18} Siegel, D. M., \& Metzger, B. D. 2018, ApJ, 858, 52

\bibitem[Stone \& Loeb(2012)]{stone12} Stone, N.,\& Loeb, A., 2012, Phys. Rev. Lett., 108, 061302.

\bibitem[Stone et al.(2013)]{stone13} Stone, N., Loeb, A., Berger, E., 2013, Physical Review D, 87, 8, 084053

\bibitem[Sun et al.(2022)]{sun22} Sun, L., Ruiz, M., Shapiro, S. L., Tsokaros, A., 2022, Physical Review D, 105, 10, 104028

\bibitem[Takahashi \& Ioka(2021)]{takahashi21} Takahashi K., Ioka K., 2021, MNRAS, 501, 5746

\bibitem[Troja et al.(2017)]{troja17} Troja E., Piro L., van Eerten H., et al., 2017, Nature, 551, 71-74 

\bibitem[Wang et al.(2018)]{wang18} Wang, J.-S., Peng, F.-K., Wu, K., \& Dai, Z.-G. 2018, ApJ, 868, 19 

\bibitem[Wang et al.(2020)]{wang20} Wang, J.-S., Peng, Z.-K., Zou, J.-H., Zhang, B.-B., Zhang, B., 2020, ApJL, 902, 2, L42

\bibitem[Willems et al.(2008)]{willems08} Willems B., Andrews J., Kalogera V., \& Belczynski K., 2008, in AIP Conf. Ser. 983, 40 Years of Pulsars, ed. C. Bassa et al. (Melville, NY: AIP), 464

\bibitem[Zhang \& M\'{e}sz\'{a}ros(2002)]{zhang02} Zhang B., M\'{e}sz\'{a}ros P., 2002, ApJ, 571, 876

\bibitem[Zhang et al.(2004)]{zhang04} Zhang B., Dai X., Lloyd-Ronning N. M., M\'{e}sz\'{a}ros P., 2004, ApJ, 601, L119

\bibitem[Zhang (2019)]{zhang19} Zhang B., 2019, Frontiers of Physics, 14, 6, 64402.

\bibitem[Zhang et al.(2022a)]{zhang22} Zhang, B.-B., Zhang, Z. J., Zou, J.-H., et al., 2022a, eprint arXiv: 2205.07670 

\bibitem[Zhang et al.(2022b)]{zhang22b} Zhang Z., Yi S.-X., Zhang S.-N., et al., 2022b, ApJL, 939, L25

\bibitem[Zhang et al.(2023)]{zhang23} Zhang Z.-J., Yin Y.-H., Wang C.-Y., et al., 2023, eprint arXiv: 2302.03215

\bibitem[Zhu et al.(2022)]{zhu22} Zhu J.-P., Wu S., Qin Y., Zhang B., Gao H., Cao Z., 2022, Astrophys. J., 928, 167


\end{thebibliography}
\end{document}